\documentstyle[12pt]{article}

\def\be{\begin{equation}}
\def\ee{\end{equation}}
\def\bea{\begin{eqnarray}}
\def\eea{\end{eqnarray}}

\newdimen\nude\newbox\chek
\def\slash#1{\setbox\chek=\hbox{$#1$}\nude=\wd\chek#1{\kern-\nude/}}
\def\QS{\slash{Q}}
\def\wpp{\bf{P}}
\def\ga{\gamma}
\catcode`\@=11
\long\def\@makefntext#1{ %\parindent 1em
\protect\noindent \hbox to 3.2pt {\hskip-.9pt
$^{{\ninerm\@thefnmark}}$\hfil}#1\hfill} %can be used

\def\thefootnote{\fnsymbol{footnote}}
 \def\@makefnmark{\hbox to 0pt{$^{\@thefnmark}$\hss}}  %original

\def\ps@myheadings{\let\@mkboth\@gobbletwo
\def\@oddhead{\hbox{} %\sl
\rightmark\hfil\ninerm\thepage}
\def\@oddfoot{}\def\@evenhead{\ninerm\thepage\hfil %\sl
\leftmark\hbox{}}\def\@evenfoot{}
\def\sectionmark##1{}\def\subsectionmark##1{}}

\begin{document}

%%%%%%%%%%%% titlepage %%%%%%%%%%%%%%%%%%%%%%%%%%%%%
\begin{titlepage}
\begin{flushright}
%{\tt hep-ph/9311xxx}\\
{BI-TP 93/62}\\
\end{flushright}
\vfill
\begin{center}
{\Large PERTURBATIVE HOT GAUGE THEORIES  - \\
RECENT RESULTS\\}
\vfill
{\large R. Baier}\\
\bigskip
{\it Fakult\"at f\"ur Physik, Universit\"at Bielefeld\\
D-33501 Bielefeld, Germany}\\
\vfill
{\large ABSTRACT}
\end{center}
\begin{quotation}
\noindent Current results in high temperature gauge theories obtained
in the context of the perturbative method of resumming hard thermal
loops are reviewed. Beyond leading order properties of the gluon
excitation, and the recent (controversial) calculations of the
damping rates are discussed. QCD predictions on plasma signatures are
exemplified by the thermal production rates of energetic
as well as soft photons.

\noindent{\sl
[Talk given at the 3rd Workshop on Thermal Field Theories and
their Applications, August 15--27, 1993, Banff, Alberta, Canada]}
\end{quotation}
\vfill
\begin{flushleft}
November 1993
\end{flushleft}
\end{titlepage}
%----------------------------PROCSLA.STY---------------------------------------
\newcommand{\symbolfootnote}{\renewcommand{\thefootnote}
        {\fnsymbol{footnote}}}
\renewcommand{\thefootnote}{\fnsymbol{footnote}}
\newcommand{\alphfootnote}
        {\setcounter{footnote}{0}
         \renewcommand{\thefootnote}{\sevenrm\alph{footnote}}}

%------------------------------------------------------------------------------
%NEW DEFINED SECTION COMMANDS
\newcounter{sectionc}\newcounter{subsectionc}\newcounter{subsubsectionc}
\renewcommand{\section}[1] {\vspace{0.6cm}\addtocounter{sectionc}{1}
\setcounter{subsectionc}{0}\setcounter{subsubsectionc}{0}\noindent
        {\bf\thesectionc. #1}\par\vspace{0.4cm}}
\renewcommand{\subsection}[1] {\vspace{0.6cm}\addtocounter{subsectionc}{1}
        \setcounter{subsubsectionc}{0}\noindent
        {\it\thesectionc.\thesubsectionc. #1}\par\vspace{0.4cm}}
\renewcommand{\subsubsection}[1]
{\vspace{0.6cm}\addtocounter{subsubsectionc}{1}
        \noindent {\rm\thesectionc.\thesubsectionc.\thesubsubsectionc.
        #1}\par\vspace{0.4cm}}
\newcommand{\nonumsection}[1] {\vspace{0.6cm}\noindent{\bf #1}
        \par\vspace{0.4cm}}

%NEW MACRO TO HANDLE APPENDICES
\newcounter{appendixc}
3~\newcounter{subappendixc}[appendixc]
\newcounter{subsubappendixc}[subappendixc]
\renewcommand{\thesubappendixc}{\Alph{appendixc}.\arabic{subappendixc}}
\renewcommand{\thesubsubappendixc}
        {\Alph{appendixc}.\arabic{subappendixc}.\arabic{subsubappendixc}}

\renewcommand{\appendix}[1] {\vspace{0.6cm}
        \refstepcounter{appendixc}
        \setcounter{figure}{0}
        \setcounter{table}{0}
        \setcounter{equation}{0}
        \renewcommand{\thefigure}{\Alph{appendixc}.\arabic{figure}}
        \renewcommand{\thetable}{\Alph{appendixc}.\arabic{table}}
        \renewcommand{\theappendixc}{\Alph{appendixc}}
        \renewcommand{\theequation}{\Alph{appendixc}.\arabic{equation}}
%       \noindent{\bf Appendix \theappendixc. #1}\par\vspace{0.4cm}}
        \noindent{\bf Appendix \theappendixc #1}\par\vspace{0.4cm}}
\newcommand{\subappendix}[1] {\vspace{0.6cm}
        \refstepcounter{subappendixc}
        \noindent{\bf Appendix \thesubappendixc. #1}\par\vspace{0.4cm}}
\newcommand{\subsubappendix}[1] {\vspace{0.6cm}
        \refstepcounter{subsubappendixc}
        \noindent{\it Appendix \thesubsubappendixc. #1}
        \par\vspace{0.4cm}}

%------------------------------------------------------------------------------
%MARCO FOR ABSTRACT BLOCK
\def\abstracts#1{{
        \centering{\begin{minipage}{30pc}\tenrm\baselineskip=12pt\noindent
        \centerline{\tenrm ABSTRACT}\vspace{0.3cm}
        \parindent=0pt #1
        \end{minipage} }\par}}

%------------------------------------------------------------------------------
%NEW MACRO FOR BIBLIOGRAPHY
\newcommand{\bibit}{\it}
\newcommand{\bibbf}{\bf}
\renewenvironment{thebibliography}[1]
        {\begin{list}{\arabic{enumi}.}
        {\usecounter{enumi}\setlength{\parsep}{0pt}
%1.25cm IS STRICTLY FOR PROCSLA.TEX ONLY
\setlength{\leftmargin 1.25cm}{\rightmargin 0pt}
%0.52cm IS FOR NEW DATA FILES
%\setlength{\leftmargin 0.52cm}{\rightmargin 0pt}
         \setlength{\itemsep}{0pt} \settowidth
        {\labelwidth}{#1.}\sloppy}}{\end{list}}

%------------------------------------------------------------------------------
%FOLLOWING THREE COMMANDS ARE FOR 'LIST' COMMAND.
\topsep=0in\parsep=0in\itemsep=0in
\parindent=1.5pc

%LIST ENVIRONMENTS
\newcounter{itemlistc}
\newcounter{romanlistc}
\newcounter{alphlistc}
\newcounter{arabiclistc}
\newenvironment{itemlist}
        {\setcounter{itemlistc}{0}
         \begin{list}{$\bullet$}
        {\usecounter{itemlistc}
         \setlength{\parsep}{0pt}
         \setlength{\itemsep}{0pt}}}{\end{list}}

\newenvironment{romanlist}
        {\setcounter{romanlistc}{0}
         \begin{list}{$($\roman{romanlistc}$)$}
        {\usecounter{romanlistc}
         \setlength{\parsep}{0pt}
         \setlength{\itemsep}{0pt}}}{\end{list}}

\newenvironment{alphlist}
        {\setcounter{alphlistc}{0}
         \begin{list}{$($\alph{alphlistc}$)$}
        {\usecounter{alphlistc}
         \setlength{\parsep}{0pt}
         \setlength{\itemsep}{0pt}}}{\end{list}}

\newenvironment{arabiclist}
        {\setcounter{arabiclistc}{0}
         \begin{list}{\arabic{arabiclistc}}
        {\usecounter{arabiclistc}
         \setlength{\parsep}{0pt}
         \setlength{\itemsep}{0pt}}}{\end{list}}

%------------------------------------------------------------------------------
%FIGURE CAPTION
\newcommand{\fcaption}[1]{
        \refstepcounter{figure}
        \setbox\@tempboxa = \hbox{\tenrm Fig.~\thefigure. #1}
        \ifdim \wd\@tempboxa > 6in
           {\begin{center}
        \parbox{6in}{\tenrm\baselineskip=12pt Fig.~\thefigure. #1 }
            \end{center}}
        \else
             {\begin{center}
             {\tenrm Fig.~\thefigure. #1}
              \end{center}}
        \fi}

%TABLE CAPTION
\newcommand{\tcaption}[1]{
        \refstepcounter{table}
        \setbox\@tempboxa = \hbox{\tenrm Table~\thetable. #1}
        \ifdim \wd\@tempboxa > 6in
           {\begin{center}
        \parbox{6in}{\tenrm\baselineskip=12pt Table~\thetable. #1 }
            \end{center}}
        \else
             {\begin{center}
             {\tenrm Table~\thetable. #1}
              \end{center}}
        \fi}

%------------------------------------------------------------------------------
%ACKNOWLEDGEMENT: this portion is from John Hershberger
\def\@citex[#1]#2{\if@filesw\immediate\write\@auxout
        {\string\citation{#2}}\fi
\def\@citea{}\@cite{\@for\@citeb:=#2\do
        {\@citea\def\@citea{,}\@ifundefined
        {b@\@citeb}{{\bf ?}\@warning
        {Citation `\@citeb' on page \thepage \space undefined}}
        {\csname b@\@citeb\endcsname}}}{#1}}

\newif\if@cghi
\def\cite{\@cghitrue\@ifnextchar [{\@tempswatrue
        \@citex}{\@tempswafalse\@citex[]}}
\def\citelow{\@cghifalse\@ifnextchar [{\@tempswatrue
        \@citex}{\@tempswafalse\@citex[]}}
\def\@cite#1#2{{$\null^{#1}$\if@tempswa\typeout
        {IJCGA warning: optional citation argument
        ignored: `#2'} \fi}}
\newcommand{\citeup}{\cite}

%------------------------------------------------------------------------------
%FOR FNSYMBOL FOOTNOTE AND ALPH{FOOTNOTE}
\def\fnm#1{$^{\mbox{\scriptsize #1}}$}
\def\fnt#1#2{\footnotetext{\kern-.3em
        {$^{\mbox{\sevenrm #1}}$}{#2}}}

%------------------------------------------------------------------------------
\font\twelvebf=cmbx10 scaled\magstep 1
\font\twelverm=cmr10 scaled\magstep 1
\font\twelveit=cmti10 scaled\magstep 1
\font\elevenbfit=cmbxti10 scaled\magstephalf
\font\elevenbf=cmbx10 scaled\magstephalf
\font\elevenrm=cmr10 scaled\magstephalf
\font\elevenit=cmti10 scaled\magstephalf
\font\bfit=cmbxti10
\font\tenbf=cmbx10
\font\tenrm=cmr10
\font\tenit=cmti10
\font\ninebf=cmbx9
\font\ninerm=cmr9
\font\nineit=cmti9
\font\eightbf=cmbx8
\font\eightrm=cmr8
\font\eightit=cmti8

%----------------------START OF DATA FILE------------------------------
%%%%%%%%%%%%%%%%%%%%%%%%%%%%%
%\begin{document}

%\begin{titlepage}
%\begin{center}
\centerline{\tenbf PERTURBATIVE HOT GAUGE THEORIES - }
\baselineskip=16pt
\centerline{\tenbf RECENT RESULTS}
\vspace{0.8cm}
\centerline{\tenrm R. BAIER}
\baselineskip=13pt
\centerline{\tenit Fakult\"at f\"ur Physik, Universit\"at Bielefeld}
\baselineskip=12pt
\centerline{\tenit D-33501 Bielefeld, Germany}
\vspace{0.9cm}
\abstracts{Current results in high temperature gauge theories obtained
in the context of the perturbative method of resumming hard thermal
loops are reviewed. Beyond leading order properties of the gluon
excitation, and the recent (controversial) calculations of the
damping rates are discussed. QCD predictions on plasma signatures are
exemplified by the thermal production rates of energetic
as well as soft photons.}
%\end{center}
%\end{titlepage}
\vspace{0.8cm}
\twelverm
\baselineskip=14pt

\section{Introduction}

At this Workshop many talks have been devoted to the properties of
systems at high temperatures and densities. In the following I will
concentrate on hot gauge theories and try to summarize some of the
recent developments and open questions in the framework of the hard
thermal loop (HTL) resummation method\cite{pia,bra/pic,fr/ta,pis}.
I mainly follow the advice by Braaten and Pisarski that
"in the end, the best way to demonstrate the consistency of the
effective expansion is to {\it apply} it to a wide range of
{\it physical processes}".

New theoretical results to be shortly discussed are on:
\begin{itemize}
\item
the gluon excitation beyond leading order;
\item
damping rates for the excitations at rest as well as the energetic
ones;
\item
the dynamical screening mechanism of mass singularities;
\item
the production of soft photons (with energies of $O (gT))$
from a quark gluon plasma and problems related to the evaluation of
their rate.
\end{itemize}

\section{Gluon self-energy: next-to-leading order corrections}

\subsection{Gluon plasma frequency}

By an impressive  calculation H. Schulz\cite{schu}
succeeded to determine beyond leading order
 the real part of the (longitudinal)
gluon self-energy in the long-wave length limit. The corresponding
solution of the dispersion relation for the gluon excitation at rest,
\be
\Omega^2 - \Pi_L (\Omega , \vec q = 0 ) = 0 ,
\ee
becomes (with $N_f = 0$ quarks and for $N$ colours) with $w \equiv Re
\Omega$ $( \equiv m_g)$,
\be w^2 = \frac{g^2 N}{9} T^2 (1 - 0.18 g \sqrt N + \ldots ),
\ee
where the corrections to the well known leading order HTL term\cite{klimov}
$  (m_g \simeq g \sqrt{N} T  /3)$  arise
from (i) hard two-loop and (ii) hard and effective soft one-loop
diagrams - after the zeroth order HTL's are subtracted.

Contrary to the original estimate\cite{bra/pic} based on power
counting arguments Schulz's explicit evaluation of 13 two-loop
diagrams (i) with hard internal momenta shows that they do not
contribute to Eq.(2) at $O (g^3 T^2)$, but only at higher order
$O (g^4 T^2)$! A clever rearrangement of the one-loop terms (ii)
simplifies their treatment. One sample of terms - the "one-loop
hard" ones in the notation of\cite{schu} - gives a negligible
contribution of $O (g^4 ln 1/g T^2)$: this, however, can only be seen
after detailed calculations as it is proportional to $g^4 NT^2 I (g
\sqrt N )$, in terms of a principal value integral
\be
I (g \sqrt N ) = \frac{1}{4\pi^2} {\bf P} \int^\infty_0
\frac{xdx}{x^2 - 1} \frac{1}{\exp (\frac{g\sqrt N}{6} x) - 1} ~ .
\ee
After expanding the Bose-Einstein factor for $g \rightarrow 0$ one
expects $I (g\sqrt N) \simeq 1/g$, i.e. an overall contribution to
Eq.(2) of $O(g^3T^2)$. However, since ${\bf P} \int\limits^\infty_0
\frac{dx}{x^2 - 1} = 0$, the integral behaves as
\be
I (g\sqrt N ) \mathop\simeq_{g\rightarrow 0} \frac{1}{8\pi^2} {\ln}~g,
\ee
and this contribution becomes negligible.

Indeed, the $O(g^3T^2)$ corrections to the plasma frequency, Eq.(2),
are only contained in the difference of the dressed minus the bare
self-energy contributions (including the tadpole diagrams),
with the one-loop momentum restricted to the
soft integration region. After a lengthy analytical evaluation the
value of the coefficient in Eq.(2) is determined numerically. It has
a negative sign, which is familiar from the behaviour of the
QCD-pressure $(N_f =0)$\cite{kap}:
\be
P \simeq \frac{(N^2 - 1)\pi^2}{45} T^4 (1 - \frac{5 g^2 N}{16\pi^2}
+ \ldots ),
\ee
i.e. indicating as Eq.(2) a phase transition for a critical value of
$g_{cr} \simeq 3.2$ for $N=3$.

Schulz' calculation shows that (i) the correction is (covariant)
gauge parameter independent in the algebraic sense\cite{ko/ku/reb},
i.e. the dependence is weighted by the on-shell condition; (ii) it
is also independent of the momentum cut-off between the
soft-and hard loop momentum; and (iii) there is consistency with the
evaluation of the damping rate in the long-wave length limit via the
dressed soft one-loop diagrams\cite{bra/pi}.

\vfill\eject

\subsection{QCD - Debye mass}

A new analysis at next-to-leading order of the static longitudinal
gluon self-energy $\Pi_{00} (w=0 , \vec q)$ and its relation to the
QCD-Debye mass is performed by A. Rebhan\cite{reb}. Here only those of
his results are presented which are of relevance in connection with
problems to be discussed in the following sections.

Instead of the usual QED-type definition of the QCD-Debye mass $m$
by $m^2~ \hat=~$ $ \Pi_{00} (w = 0 , \vec q \rightarrow 0)$,
Rebhan proposes the proper definition of $m$ by the pole position
of the longitudinal gluon propagator, i.e. by
\be
m^2 \equiv \Pi_{00} ( w = 0, \vec q^{~2} = - m^2 ).
\ee

In covariant gauges the Braaten-Pisarski resummation gives in the
effective one-loop approximation for $\vec q^{~2} \simeq - m^2$
\be
\Pi_{00} (0, \vec q ) \simeq m^2 + \delta \Pi_{00},
\ee
where
\be
\delta\Pi_{00} / g^3 T^2 \simeq c_1 ln (\vec q^{~2} + m^2 ) + c_2 +
\xi~( \vec q^{~2}+ m^2 ) I (\vec q , m).
\ee
In order to obtain a sensible value for the Debye mass $m$ the
logarithmic singularity present at the next-to-leading order has
to be "screened". Since it is due to the massless transverse gluon
propagator in the loop a (nonperturbative) "magnetic mass" $m_{mag}$
may be used as infrared cut-off.

Concerning the gauge parameter, i.e. $\xi$, dependence one observes in
Eq.(8) an algebraic on-shell independence\cite{ko/ku/reb}, in case
the
integral $I (\vec q , m)$ is well behaved at $\vec q^{~2} = - m^2$:
actually, however, it has a linear on-shell divergence. Therefore
gauge independence of $m^2$ at $O (g^3)$ requires infrared
regularisation before the on-shell limit is
performed\cite{rebb}: e.g. regularisation in $D = 3 + 2 \epsilon ,
\epsilon > 0$, dimensions leads to a $\xi$-dependence
\be
\delta \Pi_{00} \simeq g^3 T^2 \xi~( \vec q^{~2} + m^2 )^{2 \epsilon} ,
\ee
and gauge parameter independence then follows on-shell when keeping
$\epsilon > 0$.

In summary the QCD-Debye mass becomes
\be
m^2 \simeq \frac{g^2 N}{3} T^2 ( 1 + \frac{\sqrt{3N}}{2\pi}
g \ \ln \left( \frac{{\rm const}}{g}\right) + \ldots ),
\ee
when $m_{mag} \simeq g^2 T$ and $N_f = 0$ is assumed.

\section{Damping rates}

\subsection{Gluon and quark excitations at rest}

For the understanding of
 the QCD plasma properties at high temperature $T$
it is  necessary to calculate
the damping rates of the plasma excitations (waves),
the bosonic as well as the fermionic ones.
First I concentrate on the damping rates at zero momentum.
In Coulomb gauge the important results are:

\noindent
 for the
(spatially transverse and longitudinal) gluon excitation\cite{bra/pi}
\be
   \gamma_{T,L} \simeq 6.63 ~ {{g^2 N T} \over {24 \pi}} ~,
\label{gluon}
\ee
i.e. a positive constant implying the resolution of the
long standing "plasmon problem" ;

\noindent
for the  quark excitation\cite{ko/ku/ma,bra/pib}~
 (with positive and negative helicity)
\be
   \gamma_{\pm} \simeq 5.71 ~ { {g^2 C_f T} \over {16 \pi}} ~,
{}~~ for~  N_f = 3~.
\label{quark}
\ee
$C_f$ is the Casimir
for the fundamental $SU(N)$ representation.

 When calculating
 these damping constants
in  arbitrary
covariant gauges, one has to be
very careful\cite{bra/pia,bai/ku/schi,bai/ku/schib}.
In order to determine the gauge fixing, i.e.  $\xi$, dependence of the
resummed gluon and quark
 self energies, which only
comes from the gauge dependent terms in the
 resummed gluon propagator,
it is convenient to use the  Ward identities\cite{ta/wo}
for the effective propagators and vertices. It is  found that
the  gauge variations
are proportional to the corresponding inverse propagators.
E.g. the gauge variation
 of the
imaginary part of the  transverse gluon self energy is:
\be
Im~\Delta {^*\Pi_T}
(\omega ) =
 - \xi g^2 N / 2~
 q_T^4  ~ I_T (q) + ........
 \label{gaud}\ee
 \noindent where the mass-shell  is defined by
\be
q_T^2  {\strut\displaystyle\simeq \atop
{\omega\rightarrow m_g}}
 m_g~ (m_g - \omega)~. \label{mass}\ee

\noindent On shell $Im~\Delta {^*\Pi_T}$ vanishes
 provided that the corresponding integral
\be
I_T (q) = Im~ \sum_{p_0}~ \int\limits_{{\rm soft}}
 {{{d^3} p}\over {k^4 p_T^2}} ~, ~ ~ q = p + k,  \label{inta}\ee
\noindent  does not develop poles on the
mass-shell\cite{ko/ku/reb}.
However, it is singular near the mass-shell,
 \be
I_T (\omega ) \simeq
{1\over {16 \pi}} {T\over {m_g}} {1\over{(\omega -
m_g)^2}} ~,\label{intg}\ee

\noindent which implies  gauge dependence
 for the damping rate at
rest\cite{bai/ku/schi}~ (cf. Eq. (\ref{gluon})):
 \be
\delta\gamma_T
\propto \xi g^2 N T ~.
\label{dglu}\ee
\noindent
An analogous result holds for the quark
rate\cite{bai/ku/schib}.

Eq. (\ref{intg}) indicates that this
gauge dependence is related to infrared,
 i.e. to mass singularities,
 despite the fact that the damping rate itself is finite.

As already discussed in subsection 2.2
 this problem is  resolved by interchanging
limits\cite{rebb}~, i.e.
 following the prescription
 of keeping the
infrared regulator - e.g.  $\epsilon$ in dimensional
 regularisation -  different from zero
 when taking the
on mass-shell limit.

Therefore for calculating damping
rates at finite $T$ an
 infrared regulator is necessary,
  at least in covariant gauges\cite{ko/ma}.
This may be a hint for a rather complicated - gauge dependent -
 singularity structure
of propagators at finite $T$, especially near the mass-shell\cite{pi}.

\subsection{Fast moving excitations}

Recent studies of damping rates $\ga$
 of fast moving particles in hot QED or QCD
 do not yet offer satisfactory results.

As the simplest case I discuss the damping rate of a heavy fermion
of mass $M$. The
energy, the momentum and the velocity of the fermion are denoted by $E, \vec
p$ and $v$, respectively. The main interest is
 the leading order behavior, i.e. for $g\to0$
at high temperature $T$ with  $M>T$
 in  the limit
$v\to 1$.

 The difficulties arise
 from the infrared sensitive behaviour of the rate, which
 in a first approximation is
related to the Rutherford cross section and
the fermion density $n(T)$:
\be
\ga \approx n(T) \sigma_{Rutherford} \approx
T^3 g^4 \int {dq\over q^3}, \label{ruther}\ee
i.e. showing a  quadratic divergence. With an infrared cut-off of $O(gT)$
 $\ga$ becomes ``anomalous":
 its magnitude is proportional to $g^2 T$ !
Including the HTL resummation
the damping rate
-  in the one-loop approximation, in which
 the hard
energetic fermion/quark emits/absorbs one soft (dressed) boson/gluon
with momentum $q$ - may be expressed as:
\be
\ga(p_0)\simeq g^2 C_f~ T \int_{{\rm soft}} {d^3q\over(2\pi)^3} \int^{+q}_{-q}
{dq_0\over q_0} \rho_T (q_0,q) Im  G^R (p_0- q_0,\vec p-\vec q),
\label{ferm}\ee
 replacing and generalising Eq. (\ref{ruther}).
\noindent The dominant transverse
 spectral density $\rho_T(q_0,q)$ enters, with the behaviour
 for $q \to 0$~:
\be
\int^{+q}_{-q} {dq_0\over q_0} \rho_t(q_0,q)\simeq 1/q^2 ~. \label{spec}
\ee
The heavy fermion propagator is approximated
 by the (retarded) function $G^R$:
inserting  $G^R$ of a free fermion
one  finds  - instead of Eq. (\ref{ruther}) -
 a logarithmically divergent rate~:
\be
\ga(p^0\simeq E)\simeq {g^2\over4\pi} C_f~ T\int^{gT} {dq\over q}~.
\label{logd} \ee
 In QCD this integral becomes finite by introducing
 $m_{mag}\simeq g^2T$:
 evaluating $\ga$ on
the real axis $(p^0\simeq E)$ a finite value may be
  - and has been -  obtained for
quarks\cite{bra/pid,bu/ma,reba,hei/pet},
which is even gauge parameter independent\cite{na/ni/pir}.

In hot QED this logarithmic divergence  cannot
be cutoff by the  magnetic mass.
Therefore  a
 self-consistent determination of the damping rate may be attempted
as it has been first conjectured by Lebedev and Smilga\cite{leb/sm}:
 $\ga$ itself is the infrared cut-off\cite{al/pet/rio,bai/na/ni}.
The original proposal  amounts to take for  $G^R$ in Eq. (\ref{ferm})
a Lorentzian ansatz,
\be
G^R \simeq { 1 \over {p_0 - (E(p) - i\gamma)}} ~. \label{diss}
\ee

However, clarification of the ``on-shell" condition is required,
 i.e. either one keeps
$p^0$ on the real axis, $p^0\simeq E(p)$,  or
one demands self-consistency
to hold on the complex pole $p^0=E(p)-i\ga$.
Under the assumption that $G^R$ is given by  Eq. (\ref{diss})
with a complex pole  on the first (and only) sheet in the energy plane,
it turns out\cite{bai/na/ni}  that
 the infrared divergence is not screened by a
non-vanishing $\ga$! Therefore, this attempt especially fails for QED
when  Eq. (\ref{diss}) is used for the "dissipative"
 retarded fermion propagator.

For the fast QCD  damping rates Pisarski\cite{piz}
tried the self-consistent approach on the complex pole, but
 introducing in addition the magnetic mass as an
infrared cut-off, i.e. replacing the r.h.s. of Eq. (\ref{spec})
by $1 / (q^2 + m_{mag}^2)$.
 Under the condition $m_{mag} \gg \ga$,
which separates the pole from a nearby branch point,
his result is (for $v = 1)$ :
\be
\ga \simeq {g^2\over 8\pi} C_f~ T
\ln ( {m_g^2 \over { m_g^2 + 2 \ga~ m_{mag}} }) ~.
\label{pisa} \ee

\noindent However, it has been pointed out\cite{pe/pil/schi}
that this result is not stable: assuming instead of a constant
$\ga$ an energy dependent one, no self-consistent solution of
Eqs. (\ref{ferm}) and (\ref{diss}) is found for the case $v < 1$, whereas for
$v \simeq 1$ the solution is - instead of Eq. (\ref{pisa})~:
\be
\ga (E)  \simeq {g^2\over 8\pi} C_f~ T ~
   ({{ 6 \pi m_g^2} \over { E m_{mag} }})  ~.
\label{peig}\ee

Because of the decribed difficulties and inconsistencies of these
 results,
it is even  concluded that the "anomalous" damping of fast quanta
is  not observable, and thus unphysical\cite{sm}.
However, because of the  close $(?)$ connection  of $\ga$
 to transport phenomena\cite{pi,hei/pet},
especially to colour diffusion\cite{gy/se}, one has to
  argue that the simple ansatz, Eq. (\ref{diss}), for the
retarded energetic fermion propagator $G^R$ is not reflecting
realistic physical conditions.
{}From  general properties of the spectral functions, or  equivalently
from the cutting rules at finite temperature\cite{kobes/sem},
 one may deduce  that
 retarded Green  functions do not have
 complex poles on the ``physical'' sheet\cite{piz,chu/um}.
An explicit and simple example\cite{bai/ko}, in which the pole is
present only on unphysical sheets,
 shows that
 self-consistency for $\ga$ may be possible without introducing $m_{mag}$.
In QED
a fermion damping rate at leading order
\be
\ga(p^0\simeq E(p))
\simeq {e^2\over4\pi} T\ln {eT\over\ga}
\simeq {e^2\over4\pi} T\ln {1\over
e}      \label{rabai} \ee
is then derived.

\section{Production of photons from a quark - gluon plasma}

Information about the properties of the QCD plasma in its initial
stage of formation in heavy -- ion collisions is expected to be
provided by photons, real as well as virtual ones\cite{Ruus,shur/xi}.

At high temperatures
the emission of hard real photons is determined
at lowest order in the electromagnetic coupling and in $g$ by
the basic QCD processes: quark-antiquark annihilation
$( q {\bar q} \rightarrow \gamma g)$
and Compton scattering $(q g \rightarrow \gamma q) $.
However, at this order
the thermal production rates  are
 logarithmically divergent
due to massless quark exchange.
This is in contrast to the rate of thermal heavy photons,
 where the quark mass singularities
cancel. Therefore,
for real photons the mass singularities
  have to be shielded
by thermal effects  in order to derive
infrared safe predictions.

 Indeed HTL resummation provides
finite rates for {\it hard} photons
because of
 Landau damping  on the exchanged quark:
 one considers an
 effective one-loop diagram  with a quark loop with one
soft (dressed at $O(gT)$) and one hard (bare) quark line.
 No effective vertices have to be included.

The resulting emission rate for real photons with energy $E$
 is\cite{ka/li/se,bai/na/ni/re}~:
\be
E {{dW^\gamma} \over {d^3p}} \simeq {{e^2_q \alpha \alpha_s} \over
{2 \pi^2}}~ T^2~ e^{-E/T} \ln \left(
 {c \over {\alpha_s } } { E \over T } \right),
\label{pho}
\ee

\noindent where even
 the constant $c$ under the logarithm is determined,
$c \simeq 0.23$, after the hard annihilation and Compton
contributions are added.
\noindent
The emission rate is seen to be independent of the
 cut-off between soft and hard momenta. It contains, however,
a logarithmic dependence on the strong coupling constant, but
the mass singularities
 are dynamically screened.

%%%%%%%%%%%%%%%%%%%%%%%%%%%%%%%%%%%%%%%%%%%%%%%%%%%
 Other successful examples are e.g. the production
 of photon pairs\cite{bai/nie} and hard lepton pairs\cite{al/ru},
 or processes (e.g. responsible for the collisional energy loss)
 in which massless gauge bosons\cite{br/yu,br/th,tha},
 photons or gluons, are exchanged:
dynamical screening indeed cures logarithmic
mass divergencies.

For {\it soft} external photons
with energies $E$
 of $O(g T)$, or softer,
 additional  screening processes
- similar to the dilepton case\cite{bra/ay,wong} -
have to be included into the theoretical
 description\cite{bai/pe/sch,aure}:
 evaluating the dominant contribution to the
 production rate  both internal
quark propagators are assumed to be soft, and the
vertices have also to be resummed.
 As the internal quark propagators are resummed
 no divergence appears when the quark momenta
are vanishing. However, the
introduction of effective HTL vertices
leads to
unscreened collinear divergences. This may be seen from
the effective quark - photon vertex\cite{bra/pic,fr/ta}~:
\be
 ^{\ast} \Gamma^{\mu} =
 \gamma^{\mu} + m_f^2 \int {d \Omega \over 4 \pi}
{Q^{\mu} {\QS} \over \left ( Q \cdot k \right )
 \left ( Q \cdot k' \right ) }
\ , \label{vertex}
\ee
\noindent where the second
term  is the HTL
 correction -- in terms of an angular integral --  to the bare
vertex $\gamma^{\mu}$. $Q$ is a light-like vector, $Q^2 = 0$,
and $m_f^2 = 2 \pi \alpha_s T^2 /3$.
The mass singularity  arises from
the $1/ (Q \cdot k)$ factors
when both quark momenta, $k$ and
$k'$, are space-like, i.e. from:
\be
  {\wpp} \left ( {1 \over Q \cdot k} \right ) \delta (Q \cdot k') =
  \left ( {1 \over Q \cdot p} \right )
\delta (Q \cdot k') \ ,
\label{sing}
\ee
\noindent when $(Q \cdot p) = 0$ for the light-like photon
momentum   $p^2 = 0$.
Regularising this singularity
by  using dimensional regularisation of the angular integral over
 $d \Omega$ in $D =
3 + 2 \epsilon$ dimensions
the leading (singular) contribution to the soft photon production rate
 for $g \to 0$ then
reads\cite{bai/pe/sch}~:
\be
 E {dW^{\gamma} \over d^3 \vec{p}} \simeq {1 \over \epsilon}
\ { e_q^2 \ \alpha \alpha_s
\over 2 \pi^2}~ T^2 \ n_F(E) \left ( {m_f \over E} \right )^2 \ln \left ( 1
\over {\alpha_s} \right ) \ \ \ . \label{phots}
\ee
\noindent This result shows that the Braaten-Pisarski resummation does not
 yield a finite soft real photon production rate:
a logarithmic divergence remains.
 At present we do not know how to screen this mass singularity
by a consistent procedure.
Eq. (\ref{phots}) is valid for soft massless,
 i.e. non-thermalized photons:
 the quark - gluon plasma has to have a finite size, such that
 its characteristic length is smaller
than the photon's mean free path.

One may identify the diagrams which are responsible
for the singularities.
 The massless quark exchange shows up in
  two $\rightarrow$ three amplitudes, e.g. for $g^* ~ q \rightarrow
  q^* ~ g \gamma$ ($*$ denotes dressed partons), and
the singularity arises from the configuration $Q\cdot p = 0$
(cf. Eq. (\ref{sing})),
 corresponding
to a collinear singularity; here $Q$ is identified as
 the momentum carried by the
bare gluon in the HTL vertex.

 Important processes, which are sensitive to scales $\leq O(g^2 T)$,
and therefore have to be  described
 beyond the
HTL resummation, e.g.  photon bremsstrahlung from
 a QED plasma\cite{weldon}, or from a QGP\cite{go/re,cl/go/re}
 by including the effect of
 Landau-Pomeranchuk suppression\cite{lan/pom},
 and the energy loss
 due to radiation\cite{gy/pl,gy/pl/th/wa,gy/wa}
 of quarks and gluons traversing in the plasma
are under detailed  studies. However, more work is needed
in order to better  control the presently applied approximations.

\section{Conclusion}

During the last years an impressive amount of work related to the
resummation method of hard thermal loops has been performed\cite{contr},
which improves our understanding of the behaviour of gauge theories
at high temperatures. For scales from $O (gT)$ to $T$ the
resummation turns out to be in general successful in screening
fermionic and gluonic mass singularities, i.e. infrared stable
predictions are possible at leading order. One exception, however,
is the rate for the production of real photons with energies of $O
(gT)$.

Still unsolved problems remain with respect to phenomena on scales
of $O (g^2T)$ and smaller. The evaluation of the damping rates of the
(energetic) plasma excitations illustrates the present difficulties,
including the question of the existence of the magnetic screening mass.
Concerning the physical significance of these rates a detailed
description is necessary, e.g. how hard quark and gluon jets produced
during the very early stages (on time scales $\simeq 0.1 fm$) in
heavy-ion collisions may probe the dense QCD plasma subsequently
formed at times of $O (1 fm)$.

The fact that the QCD coupling constant is rather large, $g\simeq 1$,
at temperatures of the order of a few hundreds of MeV, which is the
expected realistic range for studying experimentally quark-gluon
plasma properties, requires large
extrapolations of the discussed perturbative results. Sofar they are
derived predominantly in leading, at most next-to-leading order
for high $T$ and small $g$.

\section{Acknowledgements}

Useful discussions with P. Aurenche, A. Ni\'egawa,
 S. Peign\'e , R. D. Pisarski,
A. Rebhan, K. Redlich,
 D. Schiff, H. Schulz, A. Smilga and A. H. Weldon are
gratefully acknowledged.

With pleasure I thank the organizers, especially R. Kobes and
G. Kunstatter, for the hospitality at this most stimulating Workshop.

This work was supported in part by Deutsche Forschungsgemeinschaft (DFG)
and by "Projects de Coop\'eration et d'Echange" (PROCOPE).

\end{document}